# Synthetic media and computational capitalism: towards a critical theory of artificial intelligence

David M. Berry[1]

## Abstract
This paper develops a critical theory of artificial intelligence, within a historical constellation where computational systems increasingly generate cultural content that destabilises traditional distinctions between human and machine production. Through this analysis, I introduce the concept of the algorithmic condition, a cultural moment when machine-generated work not only becomes indistinguishable from human creation but actively reshapes our understanding of ideas of authenticity. This transformation, I argue, moves beyond false consciousness towards what I call post-consciousness, where the boundaries between individual and synthetic consciousness become porous. Drawing on critical theory and extending recent work on computational ideology, I develop three key theoretical contributions, first, the concept of the Inversion to describe a new computational turn in algorithmic society; second, automimetric production as a framework for understanding emerging practices of automated value creation; and third, constellational analysis as a methodological approach for mapping the complex interplay of technical systems, cultural forms and political economic structures. Through these contributions, I argue that we need new critical methods capable of addressing both the technical specificity of AI systems and their role in restructuring forms of life under computational capitalism. The paper concludes by suggesting that critical reflexivity is needed to engage with the algorithmic condition without being subsumed by it and that it represents a growing challenge for contemporary critical theory.

This paper intervenes in current debates about artificial intelligence by introducing what I call "the Inversion," a critical concept that helps us understand a fundamental transformation in how cultural production and social consciousness are constituted under computation. Where previous analyses have focused primarily on AI's capacity to automate existing cultural processes (Alpaydin 2016; Pasquale 2020; Crawford 2021), I argue that we are witnessing a more profound shift through a qualitative transformation in which machine-generated content becomes not just indistinguishable from human production but begins to reshape the very grounds upon which we understand authenticity and experience. This change represents not merely a quantitative increase in synthetic media but a qualitative shift in how reality itself is produced and authenticated. The stakes are profound: as AI systems move from simulating to generating cultural forms, they potentially inaugurate a new regime of truth production that threatens to untether traditional mechanisms for establishing shared social reality.

This paper contributes to ongoing discussions about the human and social implications of pervasive technology (Waddell 2019; Laas 2023; Ball 2024; Placido 2024; Harris 2024). While previous work has examined specific impacts of AI systems, this analysis offers a broader theoretical framework for understanding how algorithmic systems are transforming human experience and social relations at a deeper level. I seek to explore what happens when the culture we experience and the forms of life that are predicated on this cultural context are no longer produced purely by human labour but become overwhelmed by the production of synthetic media from AI systems. This question has moved from the realm of science fiction to pressing cultural and philosophical concern.[1] As artificial intelligence systems generate increasingly convincing texts, images, and other cultural artifacts at an unprecedented scale, we potentially confront a profound reconfiguration in how meaning and human experience are constituted in societies.

✉ David M. Berry
   d.m.berry@sussex.ac.uk

[1] University of Sussex, Brighton, UK

This article contributes to key debates in contemporary scholarship (Nadin 2019; Bender et al. 2021; Coeckelbergh 2022; Mishra 2023; Vallor 2024; Gill 2024; see Coeckelbergh and Gunkel 2024). It extends critical theory's analysis of technological mediation by showing how AI systems move beyond traditional forms of technical reproduction to new modes of algorithmic genesis, posing questions about the relationship between simulation and new modes of cultural production. This happens at a moment when traditional mechanisms for establishing shared reality, from institutional authority to social trust, are already under significant strain. The proliferation of synthetic media thus threatens to accelerate what we might call a verification crisis in contemporary societies. This intervention is, therefore, urgent and advances discussions of computational capitalism by showing how this change potentially inaugurates new forms of social stratification through differential access to human experience and verification capabilities.

The theoretical contribution of this paper lies in developing new conceptual tools for understanding this transformation. Where existing approaches often treat AI-generated content as either mere simulation or straightforward automation, I argue that we need new theoretical frameworks capable of grasping how algorithmic systems are fundamentally restructuring the relationship between authenticity and artificiality. I aim to show how AI systems are not simply adding a layer of synthetic content to existing reality but transforming how reality itself is constituted and understood under computational conditions.

This extends to methodological tensions far beyond questions of research approach or academic field. As artificial intelligence and computational methods increasingly mediate both scholarly work and broader social processes, the way we understand and navigate these tensione shape not only how we produce knowledge; but what kinds of knowledge we consider valuable or possible. The challenge facing contemporary scholars lies not merely in choosing between scientific or humanistic methods, but in understanding how computational approaches might productively coexist with them or even invert them. This challenge recalls Polanyi's (2009) defence of the concept of *tacit knowledge* which he describes as forms of understanding that resist formal codification yet remain essential to effective practice. Polanyi's insight that "we can know more than we can tell" suggests fundamental limitations to purely computational approaches while also pointing toward possibilities for integration that acknowledge both explicit and implicit forms of understanding. As Berry (2014) therefore argues, the increasing prevalence of computational methods requires us to develop new theoretical frameworks that can account for both the power and limitations of digital approaches while maintaining the critical perspective essential to humanistic inquiry.

This apparent schism between computational and humanistic approaches to knowledge production, in particular, represents a significant methodological challenge in contemporary scholarship (Sartori and Bocca 2023; Romele 2023; Amoore et al. 2024). This divide manifests not merely as a technical distinction between different research methods, but as a fundamental tension in how we conceptualise and pursue understanding. The paper therefore begins by exploring what I call the *algorithmic condition*, that is, how AI and generative systems are transforming both cultural production and how we experience. It discusses the idea that machine-generated content becomes increasingly indistinguishable from human-created work, reshaping notions of cultural authenticity. Next, I examine a shift from false consciousness to what I call "post-consciousness," where the boundaries of individual and synthetic consciousness blur. I argue that this calls for new critical methods, like constellational analysis, to map the complex interplay of technology, culture, and political economy. The paper concludes by suggesting the need for resistance through *algorithmic detournement* drawing on a set of practices that redirect computational systems toward human flourishing, not just instrumental rationality. The challenge, I argue, is in developing critical reflexivity to engage with the algorithmic condition without being entirely subsumed by it.

Drawing on critical theory and extending recent work on computational ideology, I develop three key theoretical contributions: first, the concept of the *Inversion*, a cultural transformation where machine-generated works not only becomes indistinguishable from human created artifacts but also actively reshapes our understanding of "authenticity" itself; second, *automimetric production* to describe new forms of algorithmic value creation; and third, *constellational analysis* as a methodological approach for mapping the complex interplay of technical systems, cultural forms and political-economic structures under conditions of algorithmic mediation. Through these contributions, I argue that we need to use these new critical concepts to critically explore both the technical specificity of AI systems and their role in restructuring forms of life under computational capitalism. The aim is to provide a new theoretical framework for understanding how AI is transforming not just cultural production but the very possibility of critical consciousness itself. Having outlined the theoretical stakes of the argument, I will now examine in detail the algorithmic condition, exploring how it represents a qualitative break from previous forms of technological mediation while being situated within a longer historical trajectory of automated cultural production.

---

[1] The concept of "Dead Internet Theory" anticipates the idea of the Inversion by suggesting that much of the internet is now populated by bots, automated scripts, and AI-generated content rather than genuine human interaction (Ball 2024; Placido 2024). It claims "that the internet has been almost entirely taken over by artificial intelligence" (Tiffany 2021). This idea that the internet died in 2016 or early 2017, began on an imageboard called WizardChan, but today its proponents argue that an increasing portion of online activity is driven by non-human entities, capable of producing social media posts, comments, and other forms of engagement that mimic human behaviour (Keller 2018; Tiffany 2021). Interestingly, Dead Internet Theory tends to argue that the cyclical nature of repetitive content each year confirms the idea that the internet feels stale and dead.

## 1 The algorithmic condition

In this section I explore the notion of the algorithmic condition, to try to indicate the novelty of the contemporary constellation of ideas, technologies, social forms and political-economic structures that are particularly prominent today. The algorithmic condition, while representing a computational turn from previous forms of technological mediation, should nonetheless be situated within its social and historical context of growing use of automated cultural production and mechanical reproduction. However, where previous technological mediums were primarily concerned with the reproduction and circulation of human created cultural forms, the algorithmic condition marks a fundamental shift towards what I call algorithmic genesis, an additive moment which creates the conditions for the autonomous generation of cultural content through computational processes.

This transformation can be traced through three key historical moments. First, the emergence of mechanical reproduction in the nineteenth century, which Benjamin (2008) identified as undermining the "aura" of authentic cultural works. However, where Benjamin saw reproduction as diminishing authenticity, contemporary algorithmic systems invert this relationship, generating works that never had an "original" moment of human creation. This historical change is particularly visible in the shift from what Benjamin (2024) termed the optical unconscious, the camera's ability to reveal aspects of reality invisible to the human eye, to what we might call the algorithmic unconscious through AI systems capacity to generate cultural forms through the statistical analysis of training data. The second moment is the development of cybernetic systems in the mid-twentieth century, particularly what Wiener (2013) termed automated feedback loops, which presaged current forms of algorithmic automation but remained tethered to human-defined parameters. Thirdly, it is the rise of what Stiegler (2016) called grammatisation, this is where human cognitive and cultural processes became encoded in symbolic or technical systems. He writes,

> in the course of the nineteenth century, technologies for grammatising *audiovisual perception* appear, through which the flows of the sensory organs are discretised. All noetic, psychomotor and aesthetic functions then find themselves transformed by grammatisation processes. Considered in terms of political economy, this amounts to the fact that it is the functions of conception, production and consumption which are grammatised—and which are thereby incorporated into an apparatus devoted to the production of tertiary retentions controlled by *retentional systems* (Stiegler 2010, p. 11).

Yet the algorithmic condition represents something qualitatively different from these historical precedents. Where mechanical reproduction created copies of existing works, and grammatisation encoded human gestures and knowledge into technical forms, algorithmic systems now generate cultural content through automatic technical processes. This shift from reproduction to generation recalls Marx's (1981) analysis of the transition from formal to real subsumption under capital, now operating at the level of cultural production itself. Just as industrial capitalism moved from merely coordinating existing labour processes to transforming the nature of production, computational capitalism now moves beyond simply automating cultural *reproduction* to reconstituting the very grounds of cultural *generation*.

I argue that we are moving to a state where human cultural experience becomes mediated through, and constituted by, algorithmic mediation. Unlike previous forms of mechanical or digital mediation, the algorithmic condition is marked by systems that don't simply transmit or reproduce culture but actively generate and transform it through computational processes, operating at new scales and speeds.[2] Where grammatisation captured the technical recording and reproduction of this previous period of human gesture and cognition, we now confront what I call *diffusionisation*, a process through which cultural forms are probabilistically dissolved and reconstituted via computational diffusion processes. Through diffusionisation knowledge and cultural production becomes subject to what is called vector representation and latent space manipulation. These mathematical abstractions allow artificial intelligence systems to blend, morph and generate new cultural forms through probability distributions rather than deterministic rules or simple reproduction. This marks a profound shift from mere discretisation and encoding toward the autonomous generation of synthetic variations that have no original referent in human experience. Stiegler's analysis of contemporary grammatisation addresses computational systems that maintain relatively stable relationships between discrete symbols or representations. His concept of digital discretisation presumes technical systems that operate through identifiable units of information with deterministic relationships. However, I argue the movement from grammatisation to diffusionisation represents a more radical break in how cultural forms are produced and reproduced under algorithmic conditions.[3] For example, this transformation directly affects social experience, as people navigate an environment where their everyday life and shared references become detached from human experience which is no longer referent but is diffused and then mediated through these systems. The algorithmic condition becomes not just a technical possibility but an everyday reality, as our cultural experience is increasingly shaped by the creation of synthetic media whose origins and impacts become progressively difficult to ascertain.

---

[2] Joseph Weizenbaum (1976) anticipated elements of inversional logic in his observations of ELIZA, noting how users attributed understanding to simple pattern-matching programs. What he identified as a psychological tendency to anthropomorphise computers has become a structural feature of algorithmic culture, where his distinction between computational decision and human choice collapses as digital systems shape cultural production, see http://findingeliza.org

Importantly, this historical trajectory should not be understood as technological determinism. Rather, following Simondon (2017), we might see it as a process of technical individuation that occurs through the complex interplay of technical systems, social forms, and political-economic structures. The algorithmic condition thus emerges not simply through technical advancement but through what Williams (2013) termed a long revolution, that is, the gradual remediation of cultural forms through their integration into and mediation through new technical systems and social relations.

The algorithmic condition thereby alters mediation itself, moving beyond what Silverstone (1999) termed double articulation, where media technologies operate both as material objects and symbolic systems towards algorithmic articulation of diffused content and knowledge. This is where computational systems actively generate and reconstruct a novel symbolic realm itself. Traditional media primarily translated between different symbolic systems; however, algorithmic systems increasingly generate new symbolic content through automatic processes, representing not simply a quantitative increase in mediational capacity but a qualitative shift in how meaning is produced and circulated and even the kind of meanings that are generated in society. A good example of this is the rise of "AI Slop," that is, low-quality AI writing and image/video production that is pasted into social media and onto the internet in blogs and newsletter. This, often unedited, content is nonetheless almost intelligible, offering partial meaning but usually garbled through the noise generated through diffusional technologies. Meaningless content and writing thereby becomes part of our culture, particularly on social media, which we nonetheless try to understand or fit into our existing cultural horizon.

This transformation in mediation also has implications for how we understand social consciousness and social production. Drawing on Couldry and Hepp's (2016) work on what they call "deep mediatization", we might say that the algorithmic condition represents a form of deep computation where algorithmic processes don't simply mediate existing social relations but actively reconstruct them through computational means. This requires us to move beyond traditional models of mediation that assume a clear translation between human agents and technical mediators, towards approaches that can analyse the complex interplay between technical systems, social forms, and political economic structures under conditions of algorithmic mediation. This historical analysis of the algorithmic leads us to further examine the Inversion, as a critical concept that can help us understand how AI systems are deterritorialising the relationship between human and machine.

## 2 The inversion as critical concept

The concept of inversion has recently emerged in technical circles and can, perhaps, help us start to think about this phenomenon by translating it into a critical concept. Originally identified by YouTube engineers in 2013 when AI bot traffic reached parity with human traffic, for them the inversion represented a critical threshold where automated systems might begin treating algorithmic behaviour as "real" and human behaviour as "fake". As Keller (2018) described, "YouTube had as much traffic from bots masquerading as people as it did from real human visitors, according to the company. Some employees feared this would cause the fraud detection system to flip," a situation they described as an "inversion." However, beyond this technical description, I argue that the Inversion can be understood as a broader philosophical and cultural concept that can be deployed to diagnose a new social pathology emerging from how reality and artificiality are constituted in computational societies.[4]

---

[3] The emergence of AI image generation systems like Midjourney and Stable Diffusion exemplifies this shift in creative labour. Where illustrators and graphic designers previously relied on human skill and training, their work increasingly involves curating AI-generated outputs through prompt engineering. This shift from direct creation to algorithmic curation represents more than automation, rather it fundamentally transforms creative practice under algorithmic capitalism, potentially deskilling traditional creative labour while reshaping professional identities.

[4] I use "inversion," lower case to indicate the technical and musical notion, and "Inversion," capitalised, to indicate the concept.

The concept of musical inversion offers a productive metaphor for understanding the notion of algorithmic Inversion in digital culture.[5] In music theory, when we invert a chord, we maintain its fundamental identity whilst reorganising its structure through systematic repositioning of the notes. Bach's *Prelude in C Major*, which opens *The Well-Tempered Clavier*, provides a good demonstration of musical inversion. In its opening sequence, a C major chord is systematically inverted: first we hear it in root position (C–E–G), then first inversion (E–G–C), and finally second inversion (G–C–E). While each inversion remains recognisably a C major chord, each position creates distinctly different harmonic relationships and acoustic properties.

This musical metaphor can help explain how algorithmic systems maintain recognisable cultural forms while reorganising their mode of production, such that the underlying identity remains constant even as its structural relationships are transformed. The example of second inversion creating structural instability while maintaining harmonic identity is particularly relevant here as it gestures to how algorithmic systems might generate a similar form of cultural instability whilst preserving recognisable forms, suggesting that something can appear familiar on the surface but structurally destabilised through computational transformation.[6] This cultural Inversion operates across multiple domains of symbolic production, showing how algorithmic systems are deconstructing the cultural and political economic relationships between humans and society. This social understanding of Inversion helps gesture towards its impacts across different domains of cultural production. The convergence of new inversions techniques, including continuous vector representation, controlled stochasticity, and bounded novelty, creates a computational paradigm that I argue is sufficiently distinct from previous approaches to justify describing it as a qualitative break.

An example of this is the emergence of *automimetric production* on streaming platforms.[7] These are systems of "artificial streaming" where both cultural production and consumption are algorithmically automated, creating closed circuits of value extraction. Musicians and entrepreneurs have begun exploiting the political economy of streaming by creating autonomous systems that both generate and consume music, exemplifying a short circuit of cultural transindividuation where human aesthetic experience becomes incidental to the process of economic value creation. To do this they use closed algorithmic feedback loops that generate revenue. To create these feedback loops, they deploy bots to create endless variations of ambient or functional music through algorithmic composition, then use networks of automated listeners, essentially fake users, to stream these tracks continuously, generating micropayments from platforms like Spotify. In 2021, it was estimated that up to 10% of all streams might be generated by such artificial listening patterns (King 2023). This represents an interesting concrete example of a synthetic audience where the music is algorithmically generated, "listened to" by bots, and with the entire system designed to extract value from the platform's payment infrastructure rather than provide a human aesthetic experience.

What makes this case particularly revealing is how it inverts traditional models of cultural production and consumption as the "audience" is synthetic, the "creator" is increasingly algorithmic, and the actual human listeners become almost incidental to the value extraction process. This automated creation-consumption cycle demonstrates how the Inversion might restructure the cultural industries, creating what we might call *algorithmic value circuits* that operate seemingly independent of human cultural experience while still generating real economic value.[8]

---

[5] I am grateful to Prof. Ed Hughes for his comments on the music theory section discussed in this article.

[6] Similarly, in classical counterpoint, melodic inversion involves systematically transforming intervals into their opposites, where an ascending perfect fifth becomes a descending perfect fifth, maintaining mathematical relationships while creating mirror images. This suggests that the mathematical logic of musical inversion, what Schoenberg (2008) terms *developing variation*, might provide a model for understanding how algorithmic systems develop and transform cultural forms through computational processes. In his *Piano Piece Op. 19, No. 6*, for example, a three-note motif undergoes systematic transformation whilst maintaining its foundational identity. Just as Schoenberg's motif is continuously varied through inversion while remaining recognisably related to its original form, digital systems generate cultural artifacts that preserve structural relationships to their training data while producing novel variations. In addition, contrapuntal inversion demonstrates how multiple independent voices can maintain their relationships while being systematically transformed. For example, in Bach's *Contrapunctus XII* from *The Art of Fugue (BWV 1080)*, when the four-voice texture is inverted, each voice maintains its individual identity while their collective relationship is transformed through inversions, creating what is effectively a mirror image that remains musically coherent. This, I suggest, offers a useful metaphor for understanding how algorithmic systems maintain recognisable cultural forms while inverting the relationship between human and machine agency in cultural production. Just as musical inversion operates through precise mathematical relationships that transform music while preserving it, computational processes generate cultural forms that maintain surface familiarity while emerging from radically different processes.

[7] Automimetric production is exemplified by algorithmic content farms on social media, where AI systems generate synthetic influencer content and automated engagement. Meta's removal of 1.6 billion fake accounts in early 2022 demonstrates the scale of these operations (Meta 2024). These systems demonstrate a closed circuit of algorithmic production and consumption, generating economic value while potentially overwhelming human-generated content in the platform economy.

[8] Automimetric production potentially extends beyond Marx's (1992) concept of alienation, towards a form of *recursive alienation* where

A second manifestation can be found in large language models' ability to generate academic writing about themselves. When OpenAI's ChatGPT can generate seemingly convincing academic papers about itself or DALL-E can create infinite variations of artistic styles, we confront not simply a misrepresentation of reality but its algorithmic reconstruction at the level of social practice (see Thunström 2022; Kirschenbaum 2023). The Inversion thus can be seen to represent an intensification of false consciousness, what we might call *post-consciousness*. Where false consciousness involves the mystification of real social relations and algorithmic consciousness describes the mediation of experience through computational systems, post-consciousness, in contrast, marks a qualitatively new state where the very distinction between individual and synthetic consciousness becomes blurred.

I argue that this is not simply a matter of being unable to distinguish between human and machine-generated content, but rather a fundamental transformation in how consciousness itself is constituted under algorithmic conditions. Under post-consciousness, the subject doesn't merely misrecognise reality (false consciousness) or have their perception mediated by algorithms (algorithmic consciousness) but experiences a form of consciousness that is itself partly synthetic, shaped by continuous interaction with and exposure to algorithmically-generated cultural forms. This represents what we might call the third-order simulation, over and beyond Baudrillard's (1994) simulacra, where not just reality but consciousness itself becomes subject to algorithmic generation and manipulation. This form of consciousness doesn't just obscure social relations but reconstitutes our forms of life through algorithmic processes that blur the boundaries between human and machine experience. We see examples of this post-consciousness in the trance-like addiction to social media platforms, like TikTok, and the rise of "fake news" and skepticism about truth regimes prior to computation.

Perhaps most significantly, as discussed above, we can observe the Inversion represented in synthetic audiences. This can be seen as the antithesis of Dallas Smythe's concept of the audience commodity, where, instead, automated systems might generate, circulate and consume content in vast algorithmic feedback loops, often with minimal human involvement – there is no audience beyond technical media itself. Meta's own transparency reports indicate that in 2022, their automated systems were removing over 1.6 billion fake accounts quarterly, suggesting an environment where synthetic engagement potentially exceeds human interaction (Meta 2024). The metrics and measurements used to understand audience behaviour might thereby become predominantly synthetic and distorted, forcing cultural producers to adapt to algorithmic patterns of engagement to maintain their presence on a platform – in other words they would increasingly shape their content for algorithms. The Inversion is useful, therefore, for identifying a shift in how capitalism further subsumes human cultural production under computational logics.

The Inversion can be understood as marking both a technical and societal regime change, a change in ways of seeing, in which "everything that once seemed definitively and unquestionably real now seems slightly fake; everything that once seemed slightly fake now has the power and presence of the real" (Read 2018). This finds concrete expression in how different communities encounter and navigate synthetic media and are shaped by existing social and economic conditions. Rural communities might encounter synthetic media primarily through social platforms, where AI-generated content can rapidly spread misinformation about local issues, particularly in areas where local journalism has declined. Urban creative workers face different challenges, as artists and designers might confront AI systems that replicate their style and output, leading to new forms of precarity. Professionals might then need to develop new practices with journalists and academics creating verification methods, medical professionals addressing AI-shaped health misconceptions, and legal workers navigating difficult questions of veracity and evidence. Marginalized communities might bear a disproportionate burden, perhaps lacking access to digital verification tools while being particularly vulnerable to synthetic media manipulation. However, these same communities can sometimes develop collective practices for identifying and countering fake or distorted media, demonstrating how the Inversion might also spark new forms of resistance.

Having examined how the Inversion might be present itself in different domains and across communities, I now want to examine its ideological implications. The shift in analysis from technical aspects to ideological structures connects algorithmic systems to the reshaping not just cultural production but, perhaps, consciousness itself.

## 3 Computational ideology and post-consciousness

In this section I look at the notion of "computational ideology," that is, for example, the tendency to see computation as an independent force shaping social life rather than as human-created infrastructure (Berry 2014, p. 4). This reification of computational processes leads to what Marx

---

Footnote 8 (continued)

humans become estranged not just from their labour but also from cultural production itself. Unlike Adorno and Horkheimer's (2002) pseudo-individuation, this alienation is infrastructural, creating what Jaeggi (2016) might recognise as a "relation of relationlessnes" with culture itself, a structural disconnection from non-synthetic cultural experience.

identified as commodity fetishism, where social relations between humans appear as autonomous relations between things.[9] In the case of AI systems, this fetishism operates through a form of *mathematical romanticism*.[10] This is an unstable fusion of formal mathematical logic with organic, developmental narratives about machine intelligence. This romantic conception masks the material conditions of AI production while simultaneously attributing to it an almost mystical generative power.[11] This romantic conception echoes the German Romantic movement's attempt to reconcile mechanical materialism with organic vitalism, particularly visible in Novalis's search for a universal magical idealism. Just as the Romantics sought to infuse mechanical nature with spiritual life, contemporary discourse around AI sometimes seems to attribute quasi-organic properties to mathematical systems by speaking of neural networks that "learn," language models that "understand," and algorithms that "create." This becomes a form of *romantic computation* which can be understood as the dream of computer science that describes both its mechanical precision and seemingly organic becoming. The key difference is that where the Romantics sought to spiritualise mathematics (see Jahnke 1991), contemporary romantic computation computationalises spirit itself, inverting the relationship between the mechanical and the organic.[12]

While inversional logic bears striking similarities to Marx's concept of false consciousness, I argue that it is operating at a deeper pre-conscious level – affecting what we might call "control beliefs" that shape fundamental conscious thought. Where false consciousness describes how dominant ideologies obscure the real conditions of social relations, the Inversion generates an algorithmic consciousness that not only mystifies these conditions but actively reconstructs them in real-time (see also Hunter 2024). Drawing on Jaeggi's (2018) critical theory of forms of life, understood as bundles of practices, patterns of action and interpretations, we can see how the Inversion might thereby penetrate into the very fabric of everyday existence through real-time feedback mediated through digital devices.[13] The implications of this inversional logic extend beyond technical systems into the maintenance of computational forms of life, that is, forms of life as the patterns of social practice and interpretation that structure human experience. Under traditional false consciousness, workers misrecognise their real interests through ideological mystification obtained through education, social class and social structures, whereas under the Inversion, the capacity to distinguish between real and synthetic experience becomes algorithmically mediated and the patterns of interpretation and action that constitute our forms of life become framed through augmented or automated computational modes of thought.

---

[9] Lovink's (2024) *platopticon* concept, drawn from Vladen Joler's work, merges Plato's cave with Bentham's panopticon, to indicate how algorithmic governance masks itself behind voluntary social connectivity – but where we still believe we are "working for the greater good of enlightenment and progress". This also creates a form of platform fetishism, where social relations appear as autonomous algorithmic processes rather than human interactions, showing how computational ideology operates through contemporary platform infrastructures which appear to grant us new "freedoms".

[10] The concept of mathematical romanticism to capture a distinctive ideological formation in computational culture that fuses two seemingly contradictory tendencies (see Babich 2023 for a discussion of Heidegger and mathematics). On one hand, it embraces the formal rationality and axiomatic certainty of mathematical thinking, particularly in its approach to knowledge representation and algorithmic processing. On the other, it exhibits distinctly romantic characteristics in its attribution of organic, emergent, and even vital qualities to computational systems. This unstable synthesis can be seen in how AI systems are discussed, simultaneously as precise mathematical models and as quasi-organic entities that can "learn", "create", and "evolve". This paradoxical fusion bears striking similarities to the German Romantic movement's attempt to reject mechanical materialism with organic vitalism, but now manifested in computational form.

[11] Mathematical romanticism functions ideologically by obscuring the material infrastructure and human labour that underlies computational systems while promoting a mystified view of algorithmic agency. Under the Inversion, I argue that this ideology becomes particularly powerful as it helps normalise the increasing autonomy of computational systems in cultural production. When developers describe language models as "emergent" or having "creativity," they exemplify this mathematical romantic ideology that simultaneously emphasises technical precision while attributing almost magical generative powers to algorithms.

[12] The notion of *romantic computation* helps us understand a key aspect of the Inversion. Where Novalis and the German Romantics sought to discover the spiritual in the mathematical through what we might call "romantic mathematics," romantic computation inverts this relationship by attempting to discover (or perhaps produce) the mathematical in the spiritual. This is visible in how contemporary AI discourse oscillates between technical precision ("transformer models," "attention mechanisms," and "parameter spaces") and organic metaphors ("neural" networks," deep learning," and "artificial intelligence"). This instability isn't merely linguistic but might reflect a deeper inversional logic where computation is simultaneously presented as both purely mathematical and mysteriously vital. The shift from romantic mathematics to romantic computation thus marks not just a technical development but a change in how we conceptualise the relationship between mechanism and organism, calculation and creation, machine and spirit. This is a system of thought where logical and computational truths are viewed as "concrescences" (to use Whitehead's term), that is as actual occasions of computational becoming that achieve definite form through a process of creative synthesis – what we might describe as *morphic* or *generative* formalism.

[13] Jaeggi's (2018) concept of *forms of life* offers a useful theoretical framework for understanding how the Inversion operates at the level of everyday practice. For Jaeggi, forms of life are not merely cultural patterns or ways of living, but inertial bundles of social practices that represent historically developed solutions to societal problems. They are, in her terms, both normative and transformable, that is, resistant to change yet subject to rational critique and revision through immanent criticism. This approach is particularly valuable for analysing the Inversion because it helps us understand how algorithmic systems don't simply produce cultural artifacts but potentially transform the very practices through which we solve societal problems and make our world intelligible. When Jaeggi argues that forms of life are instances of problem-solving that can succeed or fail, she provides a way to critically evaluate how the algorithmic reconstruction of culture might enable or constrain human flourishing. Moreover, the emphasis on the bundles of practices that constitute forms of life helps us see how the Inversion operates not just at the level of individual artifacts or experiences but through new patterns of social practice and interpretation.

The psychological implications of the Inversion therefore extend beyond forms of alienation or false consciousness, producing instead what we might call an *inversional subjectivity*. Under platform capitalism, the subject experiences what Lovink (2024) describes as mass loneliness, which is not simply isolation but a new form of psychological existence where, paradoxically, constant connectivity produces deeper alienation. This is seen in what he terms para-social relationships and manufactured authenticity, where the subject's psychological need for recognition and connection is redirected through algorithmic mediation on social media and other digital platforms. The result is a peculiar doubling of consciousness whereby the subject is simultaneously aware of the synthetic nature of these connections while still strongly emotionally invested in them. This psychological condition recalls Weizenbaum's (1976) observations that humans readily attribute understanding and recognition to computational systems, later called the ELIZA effect, but these AI-based systems now create a situation where the very capacity for human empathy becomes destabilised as the subject of recognition itself is partially constituted by the platform – what we might call a kind of platform *severance*. The subject under the Inversion might therefore be said to exist in a constant state of what Lovink calls "optionalism," which he describes as a psychological condition where all relationships and experiences are perceived through a logic of restricted platform choice and algorithmic recommendations, leading to exhaustion and nihilism. This implies that the Inversion doesn't just transform social relations but reconstructs the psychological architecture of experience itself, mediated as it is through computational frameworks that structure and funnel conscious decision-making into a restricted palette of options or modes of thought.

These ideological transformations can be seen concretely in how different communities experience and navigate algorithmic conditions. Three critical moments are particularly useful to examine. These include, the reconstruction of social trust, the emergence of algorithmic stratification, and the digitalisation of collective memory. It is likely that the Inversion will reshape mechanisms of social trust and verification. Where previous social systems relied on what Giddens (1991) termed "expert systems" and institutional authority to authenticate reality, the proliferation of synthetic media and AI-generated content creates verification crises. Traditional markers of authenticity, such as institutional affiliation, professional credentials, or social reputation, might become unreliable as AI systems are able to convincingly simulate these markers. This creates not merely epistemological uncertainty but social epistemic opacity, where the very mechanisms for establishing shared social reality become unstable. The implications for democratic discourse and social cohesion are concerning, as the capacity to establish common ground for public debate will likely be eroded under conditions of pervasive synthetic media.

Additionally, the Inversion may generate new forms of social stratification through differential access to digital systems. As synthetic content becomes more prevalent, the ability to verify authenticity will likely increasingly depend on access to proprietary verification systems and paid AI services. This will create social groups defined by their relative capacity to distinguish between human and synthetic content. The wealthy can afford premium services that provide enhanced verification capabilities and access to "authentic" human interaction, while others must navigate a synthetic mediascape with limited tools for authentication. This technological stratification overlays existing social inequalities, potentially deepening social divides.

Lastly, the Inversion might transform processes of collective memory formation, archiving and knowledge dissemination. As AI systems mediate cultural production and circulation, they have the capacity to reshape how societies remember and forget. Traditional processes of cultural memory formation, which relied on human selection and social negotiation of significant events and narratives, are already becoming automated through algorithmic curation and affected by synthetic content generation. This might lead to a situation where collective memory becomes progressively detached from human experience and more dependent on algorithmic processes of content generation and circulation. This could take the form of individuals seeking historical or cultural information via AI interfaces, like ChatGPT, which are able to curate and organise vast quantities of information, but are not good at distinguishing between real and fake information, moreover due to their design, are likely to invent (hallucinate) missing data for an answer to a query. I observe this is already happening with undergraduate students I teach, who would rather use an LLM than a search engine to seek an answer to a question, or even a dictionary definition. The implications for cultural continuity and social identity formation are significant, as the very mechanisms through which societies maintain historical consciousness become subject to algorithmic mediation but will need more empirical work to understand.

These transformations in social trust, stratification, and collective memory point to a broader reconstruction of social reality under algorithmic conditions. As can be seen, the Inversion doesn't simply add a layer of synthetic content to existing social relations but transforms how these relations are constituted and maintained. This suggests the need for the development of pro-social algorithms through collective practices and institutions capable of maintaining social cohesion and human connection under conditions of pervasive algorithmic mediation. Such practices might include new forms of community curated verification systems, collective authentication networks, and shared practices for maintaining human everyday life amid this increasingly synthetic content.

Moreover, these social implications reveal how the Inversion operates not just as a technical process but as what Williams (2017) termed a "structure of feeling" as a shared sense of how reality itself is constituted and experienced under algorithmic conditions. We could say that this new structure of feeling is characterised by synthetic uncertainty, a pervasive sense that any given experience or interaction might be algorithmically generated or mediated – an everyday hermeneutics of suspicion. This epistemic uncertainty doesn't simply create epistemological problems but potentially reshapes the very texture of social experience, which requires new forms of critical practice and collective resistance as a response.

These profound social and ideological impacts of the Inversion will require theoretical development and new methods. Traditional critical theory, while valuable, requires rethinking to address the unique challenges of algorithmic consciousness and synthetic media. This final section therefore develops a critical theory of artificial intelligence to attempt to address both the technical specificity of AI systems and their role in reshaping forms of life under computational capitalism.

## 4 Towards a critical theory of artificial intelligence

The methodological challenges for ideology critique under conditions of the Inversion are considerable, requiring what we might call an "inversional critique" that operates simultaneously at technical, cultural, and infrastructural levels. Traditional ideology critique sought to unveil the real conditions beneath ideological mystification, but when the very processes of cultural production and interpretation are algorithmically mediated, the distinction between surface appearance and underlying reality becomes increasingly difficult to maintain. An effective critique must therefore operate both symptomatically, by examining the traces of algorithmic mediation in cultural artifacts, and systematically, analysing the computational infrastructures that generate these artifacts. This requires developing new critical methods that can account for the double articulation of algorithmic ideology, that is, both the traditional ideological content of AI-generated cultural products and the ideological implications of their algorithmic mode of production itself. Moreover, such critique must be reflexive about its own conditions of possibility, asking: how can we be certain our own critical faculties haven't been shaped by the very inversional logic we seek to analyse? The algorithmic condition therefore requires new approaches to critique and understand the technical specificity of AI systems together with their role in transforming cultural production and social consciousness.

I argue, that this suggests the need for what I call *constellational analysis*, a method of philosophical enquiry that is by turn fragmentary and constellational (see Benjamin 2008). This approach aims to examine multiple intersecting moments of technical, social, and cultural production simultaneously while remaining attentive to how, in this case, the Inversion might be restructuring our very capacity for critical thought. The aim is not simply to unmask ideology, but to trace the complex ways in which digital systems are reconstructing the very grounds upon which ideology critique has traditionally operated.

This approach is inspired by the work of Walter Benjamin who analysed the Paris arcades as crystallisation points where multiple aspects of nineteenth-century capitalism became visible simultaneously, for example in architectural forms, commodity relations, modes of perception, and changes in a collective experience. Similarly, we might analyse algorithmic platforms as points where computational systems crystallise multiple aspects of contemporary capitalism. For instance, examining automimetric production in music streaming reveals how algorithmic modes of cultural generation, synthetic patterns of consumption, new forms of value extraction, and transformations in aesthetic experience form a sociotechnical constellation. Just as Benjamin's arcades were "dream houses of the collective," these algorithmic systems might be understood as dream spaces of the computational collective, as sites where social desires and anxieties are algorithmically processed and transformed.

These AI systems do not, however, simply mimic human cultural production they invert the traditional relationship between original and copy, genuine and simulated. This transformation recalls Benjamin's concept of the "aura," the unique aesthetic authority of an original artwork which he argued was diminished through mechanical reproduction. However, AI pushes beyond it, as these systems generate works that have no clear original, no singular moment of the original creation from which copies might derive (cf AI and Corby 2022). As Berry (2024) notes, AI systems, such as LLMs (large language models), operate through vector representation and diffusion models that enable them to blend in input context with internal representations to create variations that can only with difficulty be distinguished from intentional human cultural production.[14] The Inversion suggests we are entering an era where the default assumption is that cultural artefacts are computationally generated unless proven otherwise. This represents the danger of a systemic inversion where being AI-generated is considered more "real" than being human-created.

Here I argue that critical theory can help us understand this not just as a technical phenomenon but as part of computational capitalism. The Inversion is a symptom of an accumulation regime where value extraction increasingly operates through the capture and manipulation of human cognitive and creative capacities and the circulation of culture in new value chains. However, it is important to avoid seeing the Inversion as simply a simulation of reality. Instead, it represents a "double aspect" requiring both technical and social analysis (Berry 2024). The challenge is to develop a critical theory to map these inversional processes while maintaining space for human agency and critique. This again reinforces the need for new forms of critique and practice that can engage with computational systems while maintaining human flourishing as a central concern.

To understand the Inversion, constellational analysis would therefore need to operate across multiple registers simultaneously. Moving between examining the technical architecture of AI systems, their cultural outputs, and the political economy of their production to the forms of life they engender. This might involve, for example, analysing an AI image generator through various interconnected moments, including the training data and its historical biases, the mathematical models and their embedded assumptions, the interface design and its behavioural implications, the labour relations it transforms, and the aesthetic forms it produces. The key methodological insight is that none of these moments alone captures the ideological operation of the system. It is only in their constellation that the full scope of such an Inversion becomes visible. Moreover, this approach must be sensitive to new computational temporalities, that is, the way AI systems collapse historical time into a computational space through their training data, creating a dialectical image of cultural production under algorithmic capitalism.[15]

These social and ideological shifts therefore require new critical methods capable of addressing both technical specificity and human experience.[16] This points to the importance of the need for "explainable forms of life" (Berry 2024). For example, these could be interpretative frameworks that, like a well-constructed musical counterpoint, allow us to hear both the individual voices and their systematic transformation. Explainability and interpretability need to be retooled to develop critical reflexivity that can engage with these inversional processes while maintaining space for human agency and democratic participation in shaping technological futures.[17] We need new theoretical and practical approaches that can work with and through the Inversion without being subsumed by its logic, maintaining the possibility of making them otherwise. Using this notion of constellational analysis, I want to now briefly outline some suggestions for methods for analysing this situation.

---

[14] Diffusion models represent a significant technical development in AI-generated media. They work through a two-stage process whereby first they progressively add random noise to training images until they become pure noise, then they learn to reverse this process, reconstructing images from noise. When generating new images, they reverse the process by starting with random noise and progressively "denoise" it according to text prompts or other inputs, guided by what they learned during training. This technical process itself is a striking metaphor for the Inversion as the original image is first destroyed through mathematisation (transformed into pure noise/data) before being reconstructed synthetically. Systems like DALL-E 2 and Stable Diffusion, for example, use this technique to generate highly realistic synthetic images.

[15] By using the term *algorithmic temporality,* I seek to extend Benjamin's notion of the dialectical image into the computational age. For Benjamin, the dialectical image emerged at moments where "what has been comes together in a flash with the now to form a constellation" (Benjamin 2008). Under conditions of the Inversion, AI systems create a new form of temporal constellation through their training data by collapsing vast archives of human cultural production into mathematical latent spaces that can generate new combinations. However, this algorithmic temporality differs crucially from Benjamin's dialectical image. Where Benjamin saw the dialectical image as a moment of historical awakening that could illuminate the present, I argue that algorithmic temporality tends to *flatten* historical difference into computational similarity. For example, when DALL-E or Stable Diffusion generate images "in the style of" various historical periods, they don't preserve what Benjamin called the historical index of images, that is, their specific relationship to a particular time, but rather abstract stylistic features into *ahistorical* parameters. This creates what we might call *synthetic temporality* where historical time becomes merely another dimension in the latent space of the model – these extra dimensions AI researchers call *dark knowledge*. The challenge for inversional critique is thus to develop methods that can recognize how algorithmic systems simultaneously preserve and transform historical materials, creating dialectical images at a standstill.

[16] This creates not merely a digital divide but what could be understood as a new form of class consciousness, one structured through differential access to cognitive augmentation and computational agency. Those with wealth receive personalised AI interactions designed to enhance human capacity, while those without face standardised interfaces optimised for behavioural modification and data extraction. When connected to existing educational and economic inequalities, this threatens to create what Schecter identifies as new forms of social pathology where instrumental rationality becomes the organising principle of consciousness itself (Schecter 2010).

[17] The geopolitical dimensions of the Inversion can be seen in competing models of algorithmic sovereignty. For example, in US public–private techno-economic networks (OpenAI, Microsoft), Chinese state-directed development, and European digital sovereignty through regulation (AI Act, GDPR) and a European notion of digital humanism (see Hine and Floridi 2022; also see Novelli et al. 2024). While these approaches differ, all operate within what we might call computational *realpolitik*, where control over AI capabilities becomes central to exercising global power or protecting a trading block.

Initially, I argue that we need to further develop a method of *Inversional analysis*. This would need to be a set of methodological approaches that can help trace how computational processes simultaneously maintain and transform cultural forms. This requires developing methods that can map both the technical operations of AI systems and their cultural effects. For example, when analysing large language models, we must examine not just their technical architecture but how they transform the relationship between creativity, copy, and synthetic. This is similar to Benjamin's method of literary criticism, which sought to understand works both in their technical specificity and their broader social significance. However, where Benjamin analysed how mechanical reproduction transformed the artwork's aura, we must now analyse how algorithmic generation transforms the very possibility of exercising judgement itself, for example in distinguishing between human and machine creativity.

Next, we require what we might call *recursive critique*. These are methods that can identify and challenge how algorithmic systems reshape the very grounds upon which critique operates. When AI systems can generate and flood culture with seemingly authentic academic analysis of themselves and culture, traditional approaches to ideology critique are put into question. Instead, we need methods that can maintain critical distance whilst acknowledging how computational processes transform the possibility of critical consciousness itself. This suggests developing what Stiegler (2016) termed pharmacological approaches that can analyse technical systems as both poison and cure by understanding how they simultaneously enable and constrain critical thought and elaborate criteria for curative digital systems.

Finally, we must develop what I term *constellational digital humanities* through digital and hybrid methods that can map this complex relationship between technical systems, cultural forms, and political-economic structures under algorithmic conditions. This requires moving beyond both purely technical analysis and traditional cultural critique to understand how digital systems operate across multiple registers and scales simultaneously. For example, when analysing synthetic audiences, we must examine not just the technical mechanisms of bot networks but how they transform attention economies, reshape cultural metrics, and point towards new forms of value extraction.

## 5 Conclusions

Having suggested these methodological approaches for a *critical* digital theory, I conclude by considering their implications for resistance and human agency under algorithmic conditions. This brings us back to the central challenge identified at the outset, that is, how to maintain critical reflexivity while engaging with computational systems that increasingly shape the very grounds of consciousness and experience.

While this paper examines the transformative effects of the Inversion on culture and forms of life, emerging practices of resistance deserve attention, particularly as platforms generate what Lovink terms mass loneliness and para-social relationships (Lovink 2024). Contemporary movements like Reclaim the Tech demonstrate curative possibilities for collective challenge to algorithmic governance, while the Fediverse and self-managed networks suggest alternative infrastructures that might protect and amplify human agency. However, we must also remain attentive to how resistance itself can be recuperated within computational capital.

As described above, the issues are not simply technical but require addressing, for example, what Lovink (2024) identifies as optionalism, the reduction of social life to platform-mediated choices. This suggests the need for a form of algorithmic *detournement*, through practices that redirect computational systems toward democratic and explainable forms of life.[18] Examples might include post-screenic communities, federated social networks, perma-computing movements that emphasise repair and sustainability, and collective projects building curative and creative streams outside of platform capitalism.[19] The challenge ahead lies not just in theoretical understanding but in social action to preserve and enhance human capabilities for meaning-making, social connection, and collective action under algorithmic conditions. This requires developing not only critical frameworks, as I have attempted to do in this paper, but also developing practical approaches that put human flourishing at the centre of technological development. Yet these must be understood within a broader critique of computational ideology and its tendency to reify social relations.[20] The aim is not simply to abandon platforms and AI technologies, such as in bourgeoise disconnectionism, but to develop critical practices that maintain reflexivity while building explainable forms of life under the algorithmic condition.[21]

---

[18] This draws from the Situationist concept of *detournement*, which involved subverting and repurposing elements of capitalist culture for radical political ends.

[19] Forms of resistance to the Inversion might range from direct action to infrastructural alternatives. In terms of direct action, these might include algorithmic strikes (coordinated actions against exploitative platforms), creative protest (projecting messages on technology companies' headquarters, staging die-ins in their offices and showrooms), direct action (disrupting data centres and digital infrastructure), and policy intervention (targeting legislative hearings and regulatory bodies). These practices also aim to build coalitions between cultural workers, tech workers, unions, and social movements focused on digital rights and data justice. Equally significant are attempts to build alternative infrastructures, indeed, federated social networks like Mastodon demonstrate how decentralised systems might support human-scale communities, whilst perma-computing movements emphasise sustainable, repairable technologies that resist the acceleration of platform capitalism. What makes these resistance practices particularly significant is how they attempt to reclaim human agency by working through, rather than against, computational systems.

**Author contribution**  DMB wrote the entire manuscript text.

**Data availability**  No datasets were generated or analysed during the current study.

## Declarations

**Competing Interests**  The authors declare no competing interests.

**Open Access**  This article is licensed under a Creative Commons Attribution 4.0 International License, which permits use, sharing, adaptation, distribution and reproduction in any medium or format, as long as you give appropriate credit to the original author(s) and the source, provide a link to the Creative Commons licence, and indicate if changes were made. The images or other third party material in this article are included in the article's Creative Commons licence, unless indicated otherwise in a credit line to the material. If material is not included in the article's Creative Commons licence and your intended use is not permitted by statutory regulation or exceeds the permitted use, you will need to obtain permission directly from the copyright holder. To view a copy of this licence, visit http://creativecommons.org/licenses/by/4.0/.

---

[20] Artists like Trevor Paglen and Hito Steyerl might be understood as offering a potential artistic critique of the Inversion, creating a Brechtian *Verfremdungseffekt* for the algorithmic age. We also see emerging "algofolk" practices where communities reappropriate algorithmic outputs, drawing on techniques reminiscent of Russian Constructivists like Alexander Rodchenko to critically distort/retouch/overpaint the algorithmic text. These resistant practices exploit the Inversion's inherent instabilities, offering a helpful response to computational mysticism, such as the belief in algorithms' transcendent power to optimise futures, by asserting the irreducibility of human experience to computational logics.

[21] Universities must seek to create spaces for sustained critical thought under the algorithmic condition, spaces that resist the frenetic pace of AI innovation through slow reflection and theoretical elaboration. By drawing on the notions of explainability gestured to in this paper (but see also Berry 2024), this means fostering environments where synthetic epistemologies can be critically examined and contested. The university's traditional emphasis on contemplative distance as captured in the University of Sussex's motto "to be still and know," becomes crucial for developing reasoned critique of algorithmic systems and maintaining human capacities for critical thinking amidst accelerating technological change.